\documentstyle[sprocl]{article}

\input{epsf}
\def\Journal#1#2#3#4{{#1} {\bf #2}, #3 (#4)}

\def\NPB{{\em Nucl. Phys.} B}
\def\PLB{{\em Phys. Lett.}  B}

\def\PTP{\em Prog. Theor. Phys.}

\newcommand{\bc}{\begin{center}}
\newcommand{\ec}{\end{center}}
\newcommand{\beqn}{\begin{equation}}
\newcommand{\eeqn}{\end{equation}}
\newcommand{\barr}{\begin{eqnarray}}
\newcommand{\earr}{\end{eqnarray}}

\begin{document}

\title{Finite Temperature SD Equation for Chiral Symmetry Restoration\\
in Dual Ginzburg-Landau Theory}

\author{S.~Sasaki, H.~Suganuma and H.~Toki}

\address{Research Center for Nuclear Physics, Osaka University, 
Osaka 567, Japan}


\maketitle\abstracts{
We study the chiral phase transition at $T\neq0$ 
in the dual Ginzburg-Landau theory 
using the Schwinger-Dyson (SD) equation.
In order to solve the SD equation at $T\neq0$, we provide a 
new ansatz for the quark self-energy in the imaginary-time formalism.
The recovery of the chiral symmetry is found at
$T_{_{C}}\sim 100 {\rm MeV}$ with realistic parameters, which are set
by reproducing the values of the string tension $\sqrt{\sigma}\simeq
0.44{\rm GeV}$ and the chiral 
condensate $\langle {\bar q}q \rangle \simeq -(250{\rm MeV})^{3}$ at $T=0$.
}
The dual Ginzburg-Landau (DGL)
theory is considered as an effective theory of QCD 
based on the dual Higgs mechanism in the abelian gauge \cite{suganuma}. 
In the QCD-monopole condensed vacuum, its Lagrangian is described as
%
%
\beqn
{\cal L}_{\rm DGL}=
-{1 \over 4}{\bf f}_{\mu \nu}{\bf f}^{\mu \nu}
+{m^{2}_{B} \over 2}{\bf A}_{\mu}{ 
{\varepsilon_{\lambda}{}^{\mu \alpha \beta} 
\varepsilon^{\lambda \nu \gamma \delta} n_{\alpha} n_{\gamma}
\partial_{\beta} \partial_{\delta}}
\over {(n\cdot \partial)^{2} + {n^{2}m^{2}_{B}}}} 
 {\bf A}_{\nu} + {\bar q}(i\partial \kern -2mm / -e {\bf A} \kern -2.5mm /
 \cdot {\bf H} -m_q)q, 
\eeqn
where ${\bf f}_{\mu \nu}=\partial_{\mu}{\bf A}_{\nu} 
- \partial_{\nu}{\bf A}_{\mu}$ and $n_{\mu}$ is the direction of the 
Dirac string. (The notations are the same as those in 
Ref.2.)
In the Landau gauge, the gluon 
propagator in the QCD-monopole condensed vacuum ($m_{B}\neq0$) is given by
%
%
\beqn
D_{\mu \nu}=-{1 \over k^{2}} \left [ g_{\mu \nu} -
{k_{\mu}k_{\nu} \over k^{2}} \right ] 
+ {1 \over k^{2}} {m^{2}_{B} \over {k^{2}-m^{2}_{B}}}
{ n^{2} \over {(n\cdot k)^{2}}}
X_{\mu \nu}
\label{free}
\eeqn
with $X_{\mu \nu}={1 \over n^{2}}\varepsilon^{\lambda}{}_{\mu \alpha \beta} 
\varepsilon_{\lambda \nu \gamma \delta} n^{\alpha} n^{\gamma}
k^{\beta} k^{\delta}$.
It is noted that the double pole factor ${1 \over (n\cdot k)^{2}}$ 
in the second term, which is generated as a result of QCD-monopole 
condensation,
produces the long-range correlation between quark and anti-quark due 
to its strong infrared singularity \cite{suganuma}. 
The linear confinement potential is derived from this double pole 
within the quenched approximation in the static ${\bar q}q$ 
system \cite{suganuma}.
However, this singularity causes the infrared divergence in the 
Schwinger-Dyson (SD) equation within the rainbow approximation, 
which corresponds to the quenched approximation \cite{{suganuma},{sasaki1}}. 
In previous publications,
we conjectured that the infrared screening effect for the double pole 
should be generated by the dynamics of light quarks 
\cite{{suganuma},{sasaki1}}.

Thus, we discuss the full gluon propagator ${\cal D}_{\mu \nu}$ in terms of 
the vacuum polarization tensor $\Pi_{\mu \nu}$ as 
${\cal D}_{\mu \nu}= D_{\mu \nu}-D_{\mu \lambda}[\Pi^{\lambda \sigma}]
{\cal D}_{\sigma \nu}$.
Owing to two conditions, $X_{\mu \nu}k^{\nu}=0$ and $X_{\mu \nu}n^{\nu}=0$,
we can sum over all the polarization diagrams and obtain 
the full gluon propagator as
%
%
\barr
{\cal D}_{\mu \nu}&=&
- {1 \over {k^{2}(1+\Pi_{s})}} \left [ g_{\mu \nu} -
{k_{\mu}k_{\nu} \over k^{2}} \right ] \nonumber \\
&& \;\;\;\;
+ {1 \over {k^{2}(1+\Pi_{s})}} {{\bar m}^{2}_{B} \over {k^{2}-{\bar m}^{2}_{B}}}
{ n^{2} \over {(n\cdot k)^{2}-{{\bar m}^{2}_{B} \over {k^{2}-{\bar m}^{2}_{B}}}
n^{2}k^{2}\Pi_{s}
}}X_{\mu \nu}
\earr
with $\Pi_{\mu \nu}(k)=(k_{\mu}k_{\nu}-k^{2}g_{\mu \nu})\Pi_{s}(k^{2})$ and
${\bar m}^{2}_{B}=m^{2}_{B}/(1+\Pi_{s})$.
We find that the infrared divergence at
the quenched level is removed due to 
the vacuum polarization of dynamical quarks.
Hereafter, we introduce the infrared cutoff $a$ to the double 
pole in Eq.(\ref{free}) as ${1 \over {(n\cdot k)^{2}+a^{2}}}$ 
\cite{{suganuma},{sasaki1}}. 
We also take the angular average 
on the direction of the Dirac string \cite{{sasaki1},{umisedo}}
for the gluon propagator $D^{sc}_{\mu \nu}$ with the infrared cutoff 
$a$ as
%
%
\beqn
{\bar D}^{sc}_{\mu \nu} \equiv 
{1 \over 2\pi^{2}} \int d\Omega_{n} D^{sc}_{\mu \nu}
= - {d(k^{2}) \over k^{2}} \left [  g_{\mu \nu} -
{k_{\mu}k_{\nu} \over k^{2}} \right ]\;\;,
\eeqn
where
%
%
\beqn
d(k^{2}) = 1 + {2 \over 3}{m^{2}_{B} \over {k^{2}-m^{2}_{B}}}
\left ( 1 - {2 \over a^{2}}{{a^{2}+n^{2}k^{2}} \over
{1+\sqrt{1+n^{2}k^{2}/a^{2}}}} \right )\;\;\;.
\eeqn

The SD equation in the rainbow approximation and in the chiral limit 
is given by 
%
%
\beqn
S^{-1}_{q}(p)=p \kern -2mm / - i\int {d^{4}k \over {(2\pi)^{4}}}
{\bf Q^{2}} \gamma^{\mu}S_{q}(k)\gamma^{\nu}
{\bar D}^{sc}_{\mu \nu}(p-k) \;\;.
\label{sdeq}
\eeqn
For the quark propagator, we can take 
$S^{-1}_{q}(p)= z^{-1}(p^{2})
[p \kern -1.5mm / - M(p^{2})]$; 
$z(p^{2})$ is the wave function 
renormalization and $M(p^{2})$ is the quark self-energy.
The SD equation (\ref{sdeq}) is then decomposed into a pair of 
integral equations by taking the trace of $S^{-1}_{q}(p)$ and 
$p \kern -1.5mm / S^{-1}_{q}(p)$.
The coupled equations are obtained as 
%
%
\barr
M(p^{2}) &=&  3 z(p^{2})\int { d^{4}k \over (2\pi)^{4} }\; z(k^{2})
{{{\bf Q^{2}} M(k^{2})} \over { k^{2} + M^{2}(k^{2}) }}
{d({\tilde k^{2}}) \over {\tilde k^{2}}} 
\label{masd}
\\
z^{-1}(p^{2}) &=&1 + \int {d^{4}k \over (2\pi)^{4}} 
{{{\bf Q^{2}}z(k^{2})} \over { k^{2} + M^{2}(k^{2}) }}
{d({\tilde k^{2}}) \over {\tilde k^{2}}} \left [ {{p\cdot k} \over p^{2}}
 + {{2(p\cdot {\tilde k})
(k\cdot {\tilde k})} \over  {p^{2}\tilde k^{2}}} \right ] 
\label{resd}
\earr  
with ${\tilde k}_{\mu}\equiv p_{\mu}-k_{\mu}$ in the Euclidean metric
after the Wick rotation in the $k_{0}$-plane.
Here, the Euclidean variables are simply denoted as $p$ or $k$.
In the case of $m_{B}=0$, $z(p^{2})$ 
is equal to one for all momenta in the Landau gauge,
since the second term in the r.h.s. of Eq.(\ref{resd}) is equal to zero 
by taking the angle integration \cite{higashi}.
We took therefore an assumption as $z(p^{2})=1$ for the sake of 
simplicity in previous publications \cite{{suganuma},{sasaki1}}. 
In addition, this assumption 
corresponds to the replacement 
of $d(\tilde k^{2})$ into $d(\max\{p^{2},k^{2}\})$ 
in Eq.(\ref{resd}) \cite{umisedo}.

In this paper, we solve numerically coupled integral 
equations (\ref{masd}) and (\ref{resd}) without above assumption.
As for the gauge coupling, we adopt the hybrid type of the running 
coupling \cite{{suganuma},{sasaki1}}, 
which is reduced to the perturbative-QCD running coupling 
\cite{higashi}.
We show in Fig.1 the results on the wave function renormalization $z(p^{2})$
and the quark self-energy $M(p^{2})$ as functions of $p^{2}$.
This figure indicates that $z(p^{2})$ is almost equal 
to one in the QCD-monopole condensed vacuum ($m_{B}\neq 0$).
Therefore, the assumption as $z(p^{2}) = 1$
seems reasonable.
%
%
\begin{figure}[ht]
\centerline{\epsfxsize=2.5in 
\epsfbox{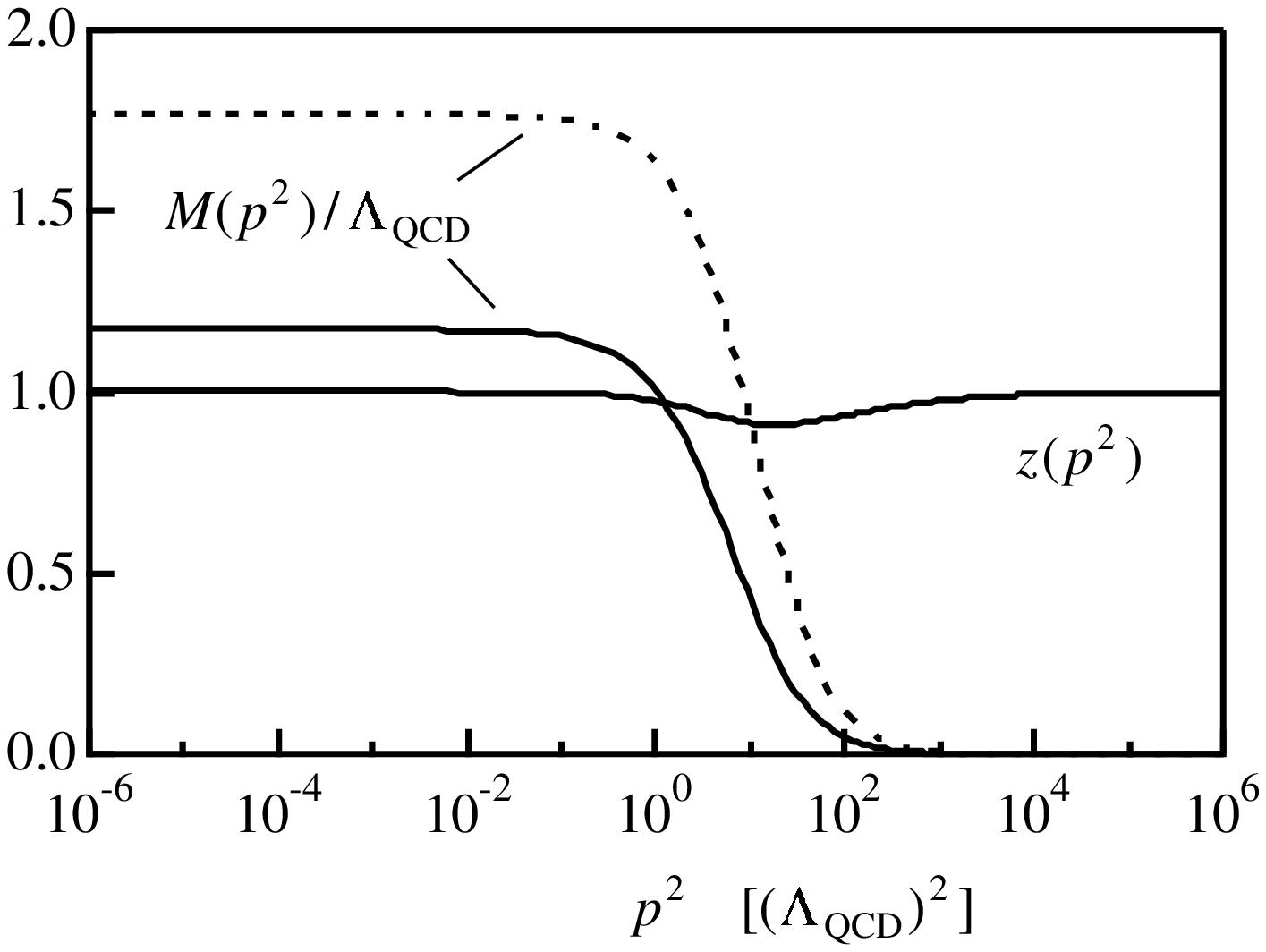}}
\vspace*{0.2cm}
{\small Fig.1~The wave function renormalization $z(p^{2})$ and the 
quark self-energy $M(p^{2})$ as functions of
$p^{2}$ with $e=5.5$, $m_B=0.5{\rm GeV}$ and $a=85{\rm MeV}$.
The result with an approximation $z(p^{2})=1$
is shown by dashed curve. The unit is taken as 
$\Lambda_{\rm QCD}\simeq 200{\rm MeV}$}
\end{figure}

Now, we formulate the finite-temperature SD equation using 
the imaginary time formalism to Eq.(\ref{masd})  
with $z(p^{2}) = 1$.
Then, the quark self-energy $M_{_{T}}(\omega_{n}, {\bf p})$
depends not only on the three dimensional momentum 
$\bf p$, but also on the Matsubara frequencies $\omega_{n}$.
As a consequence, the resulting equation is very hard to solve even
numerically.
We propose the covariant-like ansatz \cite{sasaki2}, in which a replacement 
is made for the quark self-energy at $T \neq 0$, instead of $M(p^{2}) 
\rightarrow M_{_{T}}(\omega_{n}, {\bf p})$, as 
%
%
\beqn
M(p^{2}) \rightarrow M_{_{T}}({\bf p}^{2}+\omega_{n}^{2})\equiv
M_{_{T}}(\hat p^{2})
\eeqn
with $\hat p^{2}\equiv{\bf p}^{2}+\omega_{n}^{2}$ and 
$\omega_{n}=(2n+1)\pi T$. 
It is noted that this ansatz guarantees that
the finite-temperature SD equation in the limit $T \rightarrow 0$ is 
exactly reduced to the SD equation at $T = 0$. This fact is also
confirmed by numerical calculations as shown in Fig.2.
The final form of the SD equation at $T \neq 0$ is derived as
%
%
\beqn
M_{_T}(\hat p^2) 
= {3T \over {8 \pi^2}} \sum^{\infty}_{m=-\infty}
\int^{\infty}_{\omega^2_m} {d{\hat k}^2} {\int^{1}_{-1}{du}}\; 
\sqrt{\hat k^2 - \omega^2_m}
{{\bf Q}^2M_{_T}(\hat k^2) \over {\hat k^2+M^2_{_T}(\hat k^2)}} 
{d({\tilde k^2_{nm}}) \over {\tilde k^2_{nm}}}\;,
\label{fsd}
\eeqn
where $\tilde k^2_{nm}={\hat k}^2+{\hat p}^2-2u
\sqrt{({\hat p}^2-\omega^2_n)({\hat k}^2-\omega^2_m)}
-2\omega_m\omega_n$. 
We then solve the SD equation by setting $\omega_{n}=0$ in the r.h.s. of 
Eq.(\ref{fsd}) \cite{sasaki2}.
We show in Fig.~2 the quark self-energy $M_{_T}({\hat p}^2)$ as a 
function of ${\hat p}^2$ at finite temperature.
No nontrivial solution is found in the high temperature region \cite{sasaki2}, 
$T\stackrel{>}{\scriptstyle \sim}110{\rm MeV}$.  
In other words, the chiral symmetry is restored at high temperature.

In summary, we have confirmed that the infrared singularity problem is
avoided due to the screening effect of the quark vacuum polarization.
By solving numerically the coupled SD equations at $T=0$, we have found 
that the wave function renormalization is almost equal to one in all momenta
and the quark self-energy has non-zero value.
We have solved the the SD equation at $T\neq0$
numerically with the covariant-like ansatz
and found that the chiral symmetry is restored at high temperature;
$T_{_{C}}\sim 100 {\rm MeV}$ with realistic parameters \cite{sasaki2}.
%
%
\begin{figure}[ht]
\centerline{\epsfxsize=2.5in 
\epsfbox{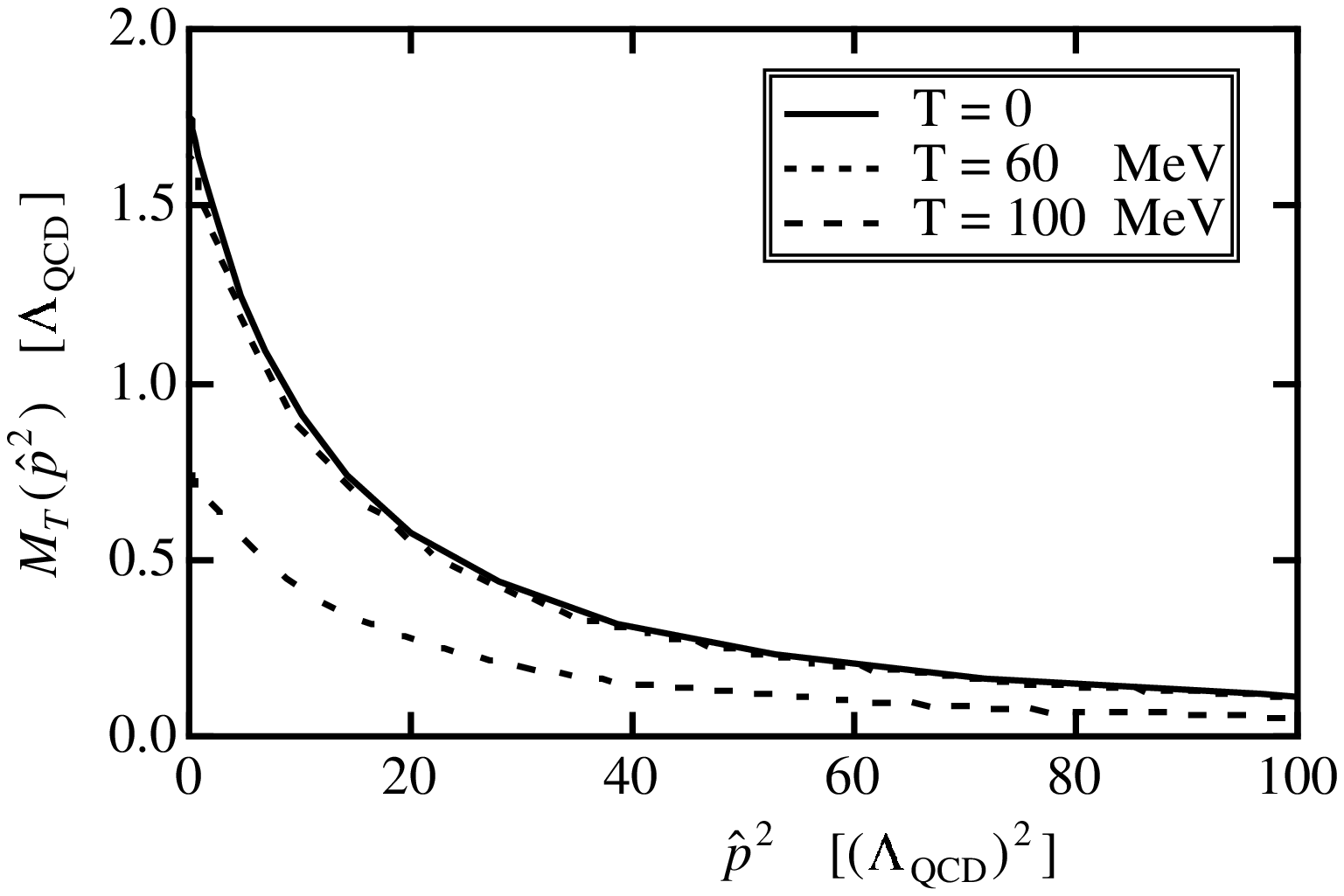}}
\vspace*{0.2cm}
{\small Fig.2~The quark self-energy $M_{_T}({\hat p}^{2})$
as the function 
of ${\hat p}^2$ at $T=0,~60$ and $100{\rm MeV}$.
The same parameters are used as in Fig.1.}
\end{figure}

\end{document}